\documentclass[12pt,a4paper,final]{iopart}

\usepackage{graphicx}
\pdfoutput=1
\usepackage{amsfonts,amsmath,amsthm,amssymb,amsopn}
\usepackage{dsfont,bbm}
\usepackage{mathrsfs}
\usepackage{leftidx}
\usepackage{physics}
\usepackage{cite}
\usepackage{booktabs}
\usepackage[retainorgcmds]{IEEEtrantools}
\usepackage{soul,xcolor}
\usepackage{color}
\definecolor{darkblue}{rgb}{0.,0.,0.5}
\definecolor{darkred}{rgb}{0.6,0.,0.}
\definecolor{darkgreen}{rgb}{0.,0.8,0.}
\definecolor{darkpink}{rgb}{1,0.08,0.5}

\usepackage[breaklinks=true,colorlinks=true,linkcolor=darkblue,urlcolor=magenta,citecolor=magenta]{hyperref}
\numberwithin{equation}{section}
\usepackage{cleveref}
\def\be{\begin{equation}}
\def\ee{\end{equation}}

\global\long\def\al{\alpha} \global\long\def\be{\beta}
\global\long\def\ga{\gamma}

\eqnobysec

\begin{document}
\title{ Universality  in many-body driven systems with an umbilic point}

\author{Johannes Schmidt}
\address{Bonacci GmbH, Robert-Koch-Str. 8, 50937 Cologne, Germany}
\author{\v Ziga Krajnik}
\address{Department of Physics, New York University, 726 Broadway, New York, NY 10003, United States}

\author{Vladislav Popkov} \address{Faculty of Mathematics and
  Physics, University of Ljubljana, Jadranska 19, SI-1000 Ljubljana,
  Slovenia} \address{Department of Physics, University of
  Wuppertal, Gaussstra\ss e 20, 42119 Wuppertal, Germany}

\begin{abstract}
We study  stationary fluctuations
of conserved slow modes in a two-lane model of hardcore particles which 
are expected to show universal behaviour.  
Specifically,  we focus on  the properties of fluctuations at a special  umbilic  point where the characteristic velocities coincide.    At large space and time scales, fluctuations are described by  a system of stochastic Burgers equations studied recently in
 \cite{2024Spohn}. Our data suggest coupling-dependent scaling functions and, even more surprisingly, coupling-dependent dynamical scaling exponents, distinct from  KPZ scaling exponent typical for surface growth processes.
\end{abstract}


\maketitle

\pagestyle{empty}
\tableofcontents
\pagestyle{headings}

\flushbottom
\clearpage

\section{Introduction}

One-dimensional driven stochastic systems of  particles interacting with a short range interactions  with one global
conservation law belong to the renowned Kardar-Parisi-Zhang (KPZ) universality class \cite{KPZ},  with 
 the exact scaling function found by Pr\"ahofer and Spohn \cite{Prae04}.  KPZ universality 
is expected under rather general assumptions which include a finite steady-state correlation length,  stationarity 
of spatially homogeneous distributions, a presence of noise and of long-lived modes stemming from the conservation
law.  Physically, realizations of KPZ universality have been demonstrated in  a series of beautiful experiments by
Takeuchi et al \cite{Takeuchi,Takeuchi2020}.
Dynamical behavior of many-body systems with  the two or more long-lived modes at large space and time scales hase been shown to be well
 described by nonlinear fluctuating hydrodynamics and mode-coupling theory \cite{Spoh14}.   
A key assumption of the resulting nonlinear fluctuating hydrodynamics and mode coupling analysis is a spatial separation of  long-lived modes with time,  meaning that velocities of the modes are all different.
Such an assumption of strict hyperbolicity is also a cornerstone of the Lax theory of shocks \cite{Lax2006}.

For strictly hyperbolic systems with several conservation laws mode-coupling theory \cite{Spoh14} predicts  dynamical scaling exponents $z_\alpha$ given by the ratio
of neighboring Fibonacci numbers $z_\alpha=2,\,3/2,\,5/3,\,8/5,\ldots, \varphi$  \cite{2015Fibonacci}, where $\varphi= (\sqrt{5}+1)/2$ is  the golden ratio. The first two members of the family correspond to diffusive and KPZ universality classes.

The presence of an umbilic point,  where the velocities of two or more modes coincide,  invalidates the standard shock picture  \cite{Lax2006} and gives gives rise to unusual dynamical properties, which appear also at large space and time scales, e.g. stable umbilic shocks, which are unstable in a strictly hyperbolic system \cite{Umbilic2012}.  
Umbilic shocks govern unusual boundary-driven phase transitions \cite{UmbilicTransitions2013}.  
Notably, the umbilic point does not require any exotic microscopic dynamic rules for its existence, and  appears naturally as  a mere   consequence of a left-right reflection symmetry of rates in bidirectional systems \cite{Umbilic2012}.

A recent paper \cite{2024Spohn} treats the question of universality of a weakly hyperbolic system with two conserved 
quantities using two approaches: firstly,  by numerical integration of a system of two stochastic Burgers equations, and secondly,  by  Monte Carlo simulation of a two-lane bidirectional particle model with an umbilic point.  
In particular, the scaling of space-time correlations in those two systems is investigated to infer the dynamical exponent and the distribution of the time-integrated currents. 
An important feature of both models is a so-called cyclicity condition for the mode-coupling matrices $G^\al$  which guarantees  time-stationarity of the Bernoulli measure.  
For both cases the authors find a match between the space-time correlators for both lattice and continuous model. 
The authors of  \cite{2024Spohn}, relying on the available direct numerical simulations, conclude that the dynamical exponents  for the umbilic modes are KPZ-like, $z_1 =z_2= 3/2$,  as in the the single-component Burgers equation,  and that the scaling functions are  coupling-dependent.  

\begin{figure*}[t!]
\includegraphics[width=0.98\linewidth]{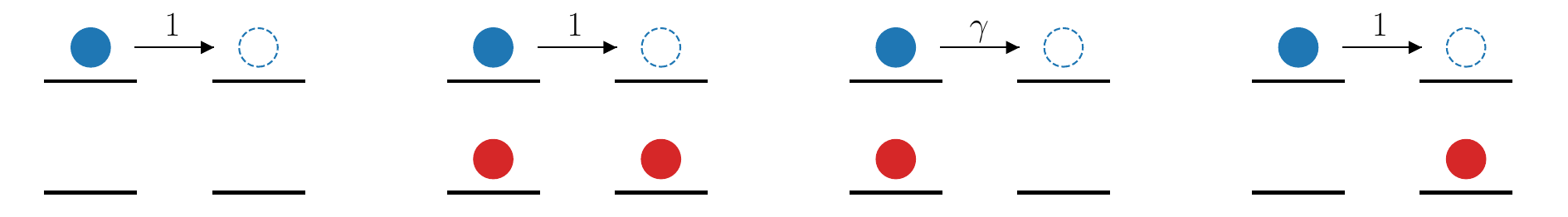}
\caption{Dynamics of the bidirectional two-chain model. Hopping rates for the right movers in the upper chain (blue circles) are normalized to unity except for $0< \gamma< \infty$ which sets the inter-chain interaction \cite{GunterSlava_StatPhys03}.  Hopping rates of the left movers in the lower chain (red circles) are obtained from those of the right movers by a left-right reflection.}
\label{Fig_BidirectionalModel}
\end{figure*}

The purpose of the present communication is to present evidence that not only the scaling functions but also   the dynamical exponents for the umbilic modes are distinct from the KPZ scaling exponents,  $z_1 , z_2 \neq  3/2$.  This contradicts one of main statements of \cite{2024Spohn}.   
Namely,  our results strongly suggest that both scaling functions and dynamical scaling exponents are coupling-dependent.  
We show that the reason for the contradiction is the short time window employed in \cite{2024Spohn},  based on which the observations and the conclusions were made.  We have performed Monte-Carlo simulations over much larger time scale $t\leq 10^5, L= 5 \times 10^5$ for the two-lane lattice model, compared to  $8000<t<16000$,  $L=2^{16}\approx 6.5 \times 10^4$ in \cite{2024Spohn}, {see \ref{app:D} for a comparison of the data. Our numerical simulation of the system of coupled Burgers equations studied in \cite{2024Spohn} similarly shows deviations from  $z_1 , z_2 \neq  3/2$ at longer times.}
In view of the above, umbilic universality appears to be an even more exciting and challenging topic which needs further studies. Indeed, our findings are in sharp contrast with the strictly hyperbolic case where a discrete family of the dynamical exponents (Fibonacci universality classes \cite{2015Fibonacci}) exhaust all possible values of the dynamical exponents of a model  with two conserved quantities.

\section{Description of an umbilic point.  Saddle and convex topologies}
\label{sec::Description of the umbilic point}
\begin{figure}[t]
\includegraphics[width=\linewidth]{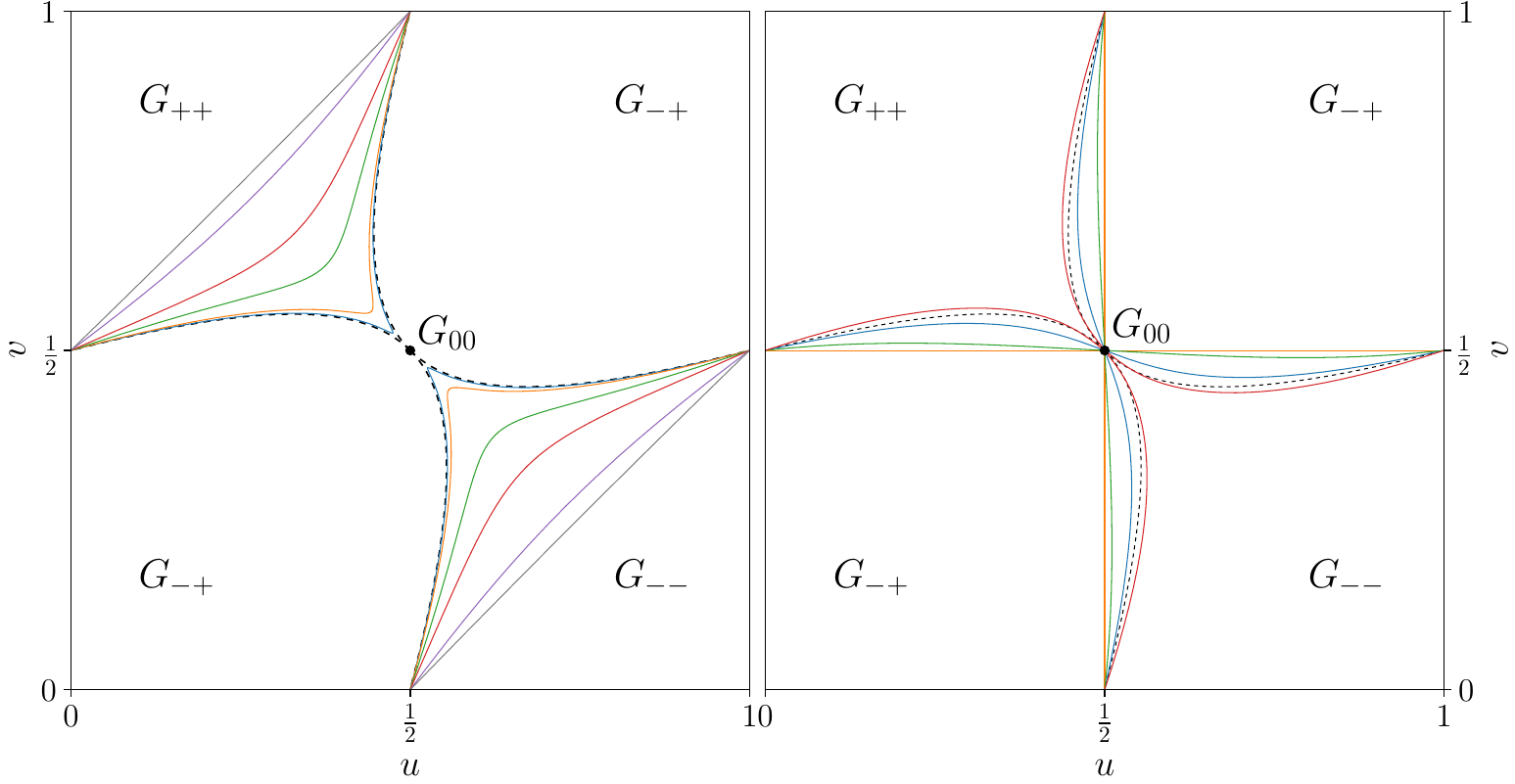}
\caption{Splitting of the physical region into domains $G_{++},G_{--},G_{-+}$
according to signs of characteristic velocities, for different $\gamma$. Boundaries
between the domains, on which one characteristic velocity vanishes
$c_{i}(u,v)=0$, are marked by colored lines. 
The point $G_{00}$ at $u=v=1/2$ (black circle) is an umbilic point where
$c_{1}=c_{2}=0$ for any value of $\gamma$. 
Region boundaries for $\gamma_{\rm crit}$ (dashed black lines) demarcate between distinct topologies of the umbilic point.
(upper panel) For $\gamma = \gamma_{\rm crit} - \Delta \gamma$ with $\Delta \gamma= 0.002, 0.01, 0.05, 0.1, 0.2, 0.25$ (blue to gray) the umbilic point is an isolated point 
(lower panel) For $\gamma = \gamma_{\rm crit} + \Delta \gamma$ with $\Delta \gamma = 0.03\, ({\rm blue}), 0.75\,  ({\rm orange}), 3\, ({\rm green}), \infty\, ({\rm red})$ the umbilic point is at the intersection of region boundaries.}
\label{FigSplitting}
\end{figure}

We consider a two-lane bidirectional model of left-moving and right-moving particles, with hardcore interaction between them (each site cannot be occupied by more than one particle) proposed in \cite{GunterSlava_StatPhys03},
a schematic representation of which is given in Fig.~\ref{Fig_BidirectionalModel}.   {
A blue particle at site $k$ in the upper lane can hop to the right neighbouring site $k+1$ provided it is empty,  with the rates shown in Fig.~\ref{Fig_BidirectionalModel},
which depend on the occupation of sites $k,k+1$ in the other (lower) lane.  
Likewise,  on the lower lane (red particles) can hop to the left,  with the hopping rates  obtained from those of the  right movers by a reflection.
The model contains one inter-chain interaction parameter $\gamma \in [0, \infty)$, 
and is a reduction of a more general two-lane exclusion process,  proposed in \cite{GunterSlava_StatPhys03}.  
 For $\gamma=1$  the system decouples into two separate totally asymmetric exclusion processes, characterized by KPZ universality.

The time evolution of the model is described by a continuous-time Markov process with above described rates.  It was shown in 
\cite{GunterSlava_StatPhys03} that the stationary state probabilities $P_{n_1,n_2,\ldots ,n_L}^{m_1,m_2,\ldots ,m_L} $ 
of configurations on a periodic lattice of length $L$ is given by
\begin{align}
P_{n_1,n_2,\ldots ,n_L}^{m_1,m_2,\ldots ,m_L} = Z^{-1} \prod_{k=1}^L e^{-\nu n_k m_k} \label{eq:ExactNESS}
\end{align}
where $e^{\nu}=\ga$ and  $n_k,m_k$ are occupation numbers $n_k,m_k=0,1$ for the lower and upper chain respectively. 
} The factorization  (\ref{eq:ExactNESS})
allows one to obtain  explicit expressions for steady microscopic currents $j_1,j_2$ as a function of the average particle densities on upper and lower lanes $0\leq u,v \leq 1$.
The eigenvalues of the flux Jacobian,
\begin{align}
{\cal D}_J(u,v)&=\left(
\begin{array}{cc}
\frac{\partial j_1}{\partial u} & \frac{\partial j_1}{\partial v}\\
 \frac{\partial j_2}{\partial u} & \frac{\partial j_2}{\partial v}
\end{array}
\right),\\
{\cal D}_J(u,v) \psi_\alpha(u,v) &= c_\alpha(u,v) \psi_\alpha(u,v),
\label{J psi=c psi}
\end{align}
$c_{1}(u,v)$ and $c_{2}(u,v)$ are real continuous functions and correspond to characteristic velocities, i.e.~velocities of local perturbations above stationary state with densities $u,v$. At $u=v=1/2$ the  flux Jacobian becomes degenerate, $c_1=c_2=0$,  yielding a unique umbilic point in the $u,v$  plane (all other $u,v$ points correspond
to $c_1 \neq c_2$).  
The phase space $0\leq u,v\leq 1$ can consequently be divided into
regions the $G_{\pm \pm}$ and $G_{-+}$ according to signs of $c_{1}$ and $c_{2}$ respectively, e.g. $G_{-+}$ is the region where $c_1<0$ while $c_2>0$. The umbilic point $G_{00}$ at the center $u=v= 1/2$ is either a point of contact 
of all three regions regions for $\gamma> \gamma_{crit}=1/4$
or an isolated point inside the $G_{-+}$ phase for $0 < \gamma< 1/4$ (in this case,  $G_{++}$ and  $G_{--}$ are separated, see Fig.~\ref{FigSplitting}).
The two cases $\gamma> \gamma_{crit}$ and $\gamma< \gamma_{crit}$ thus  correspond to distinct topologies. 
Indeed,  the surface topology of the currents  $j_\alpha(u,v)$ at the umbilic point is convex for  $ \gamma_{crit}\geq \gamma$ and forms a saddle for $\gamma< \gamma_{crit}$, see \cite{Umbilic2012}.

In case of non-zero characteristic umbilic point velocity  $c_1(u_0,v_0)=c_2(u_0,v_0)=c$,  the same reasoning  applies
after a renormalization  $c_\alpha(u,v) \rightarrow c_\alpha(u,v)-c$.
In other words, an umbilic point at $u_0,v_0$ with $c_1(u_0,v_0)=c_2(u_0,v_0)=c$ is isolated if there exists an $\epsilon_0>0$ such that for any $0<\epsilon<\epsilon_0$, and  $0\leq \varphi<2 \pi$,
$c_\alpha(u_0+\epsilon \cos\varphi ,v_0 +\epsilon \sin\varphi) \neq c$, $\alpha=1,2$.
In a system with  $K>2$ conservation laws further topological possibilities appear, since the phase space becomes  $K$-dimensional.

Ref.~\cite{2024Spohn} shows that the bidirectional model at the unique umbilic point $u=v=1/2$ and for any coupling $\gamma$  corresponds (in the framework of Taylor expansion to the second order of the stationary currents as function of particle densities)  to a system of coupled stochastic Burgers equations with cyclic coupling matrices and a degenerate flux Jacobian.
Moreover, for the critical value of the coupling $\gamma_{\rm crit} = 1/4$ (the $\gamma$ value separating the isolated and non-isolated umbilic point scenario)   the mode coupling matrices become especially simple, and correspond to a dynamics, maximally distant from a pair of decoupled scalar KPZ growth processes. 

\begin{figure}[ptb]
\includegraphics[width=0.98\columnwidth]{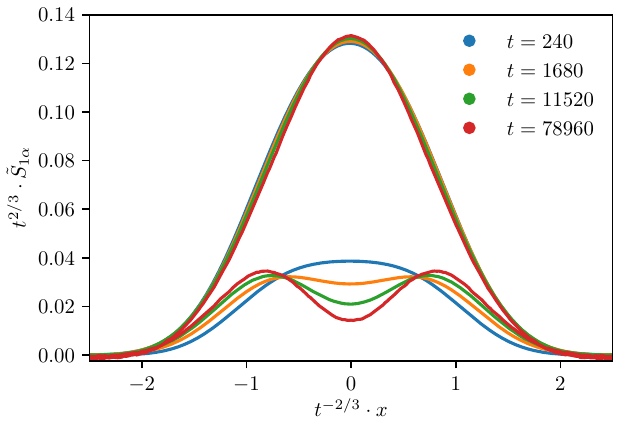}
\caption{Rescaled bare diagonal $\tilde S_{11}$ (upper curves) and off-diagonal $\tilde S_{12}$ (lower curves) correlators of the two-lane process at different times rescalied with the dynamical exponents $z_1= z_2=3/2$.
System parameters: $L=5 \times 10^5$, $\tau=80$, $M=10^4$, $P=983$, $R=1$, see \ref{app:B} for details.
}
\label{fig:unrotated_correlation}
\end{figure}

\section{Numerical investigation of the dynamical structure factor}
We first investigate the ``bare" space-time correlation matrix (dynamical structure factor) of the two-lane Markov process 
 \begin{align}
&\tilde{S}_{\alpha \alpha}(x=ka,t) = \langle \eta_k^{\alpha}(t) \eta_0^{\alpha}(0) \rangle,
\end{align}
with particle densities $\eta_x^\alpha(t)$ in Fig.~\ref{Fig_BidirectionalModel} where $\alpha$ is the lane index, $a$ the lattice spacing and $k$ the lattice position. To facilitate the comparison to other models and establish a connection with the system of coupled stochastic Burgers equations of Ref.~\cite{2024Spohn} we also consider a basis where the static covariance matrix becomes an identity matrix, see \ref{app:A}.  
In this basis the space-time correlation matrix $S_{\alpha \beta}$ becomes diagonal,  with the diagonal elements  given by 
\begin{align}
S_{11}(x,t)&=(1+\gamma^{1/2}) \left(\tilde{S}_{11}(x,t) + \tilde{S}_{22}(x,t) +\tilde{S}_{12}(x,t) + \tilde{S}_{21}(x,t)\right) \label{eq:StildeS11}\\
S_{22}(x,t)&=(1+\gamma^{-1/2}) \left(\tilde{S}_{11}(x,t) + \tilde{S}_{22}(x,t) -\tilde{S}_{12}(x,t) - \tilde{S}_{21}(x,t)\right)\label{eq:Sab}
\end{align}
Up to a rescaling, the functions $S_{\alpha \alpha}(x,t)$ are identical to those studied in  \cite{2024Spohn}. 
According to the scaling hypothesis, at large space and time scales one expects 
\begin{equation}
S_{\alpha\alpha}(x,t) \sim t^{-1/z_\alpha} f_{\alpha} \left(xt^{-1/z_\alpha}\right), 
\label{scalingform}
\end{equation}
where  $z_\alpha$ is a dynamical exponent. 
To quantify the scaling \eqref{scalingform} we have performed extensive Monte Carlo simulations for a periodic system of size $L=5\times 10^5$ up to $t=10^5$ time units, starting from spatially homogeneous states with densities $1/2$ on both lanes. Measurement were done at all sites in parallel, since $S_{\alpha\alpha}(x,t)$ depends only on the distance between two points by translational invariance. Further details on numerical simulations can be found in the \ref{app:B}. 

\begin{figure}[t]
\includegraphics[width=0.98\columnwidth]{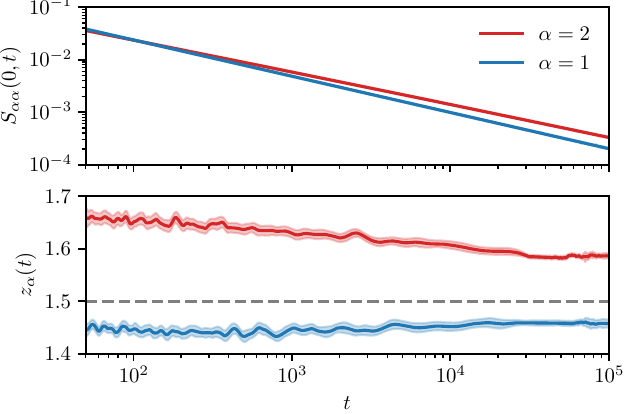}
\caption{Upper panel: Scaling of the maxima $\max S_{\alpha \alpha}(t)$ of the distribution,
$\max S_{\alpha \alpha}(t) \sim t^{-1/z_{\alpha}}$ according to scaling hypothesis. Lower panel: Dynamical exponents $z_1(t)$ and $z_2(t)$ as functions of time. 
Parameters: $L=5\times 10^5$, $\tau=0$, $M=1$, $P=2500$, $R=50$, and $p=0.01$ see \ref{app:B} for details.}
\label{FigZ(t)}
\end{figure}

In chosen states the structure factor is invariant under exchange of lanes $\tilde S_{\alpha \beta} = \tilde S_{\beta \alpha}$, leaving two independent bare correlators shown in Fig.~\ref{fig:unrotated_correlation} rescaled with $z_{1}=z_2=3/2$. While the rescaled data for the diagonal correlator $\tilde S_{11}$ nearly collapses for different times, the off-diagonal correlator $\tilde S_{12}$ manifestly do not. We instead observe a gradual deepening of the central minimum accompanied by thickening tails due to the sum rule, precluding a uniform scaling collapse.

\begin{figure}[htb]
\includegraphics[width=0.48\columnwidth]{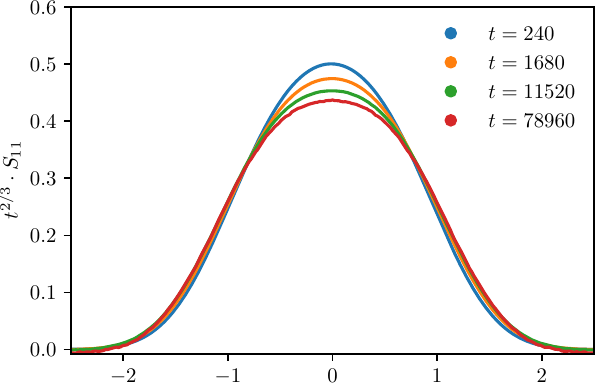}
\includegraphics[width=0.48\columnwidth]{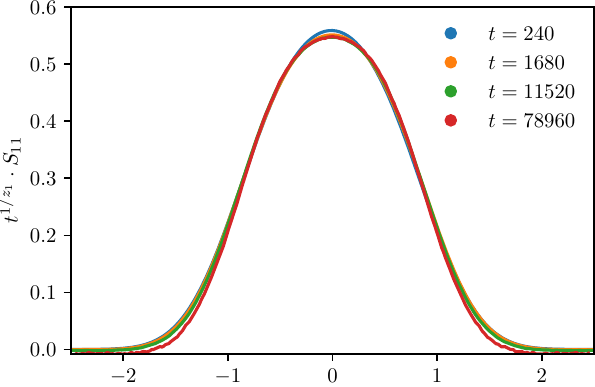}

\includegraphics[width=0.48\columnwidth]{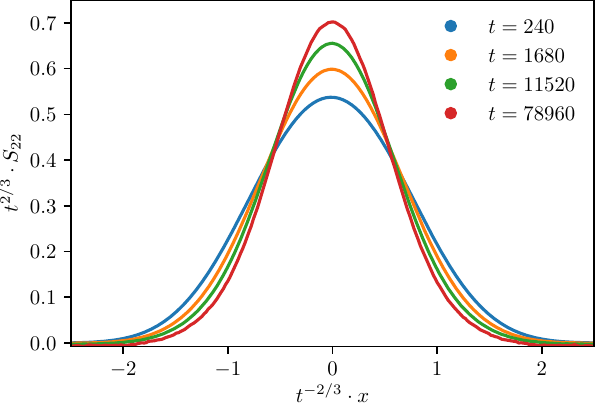}
\includegraphics[width=0.48\columnwidth]{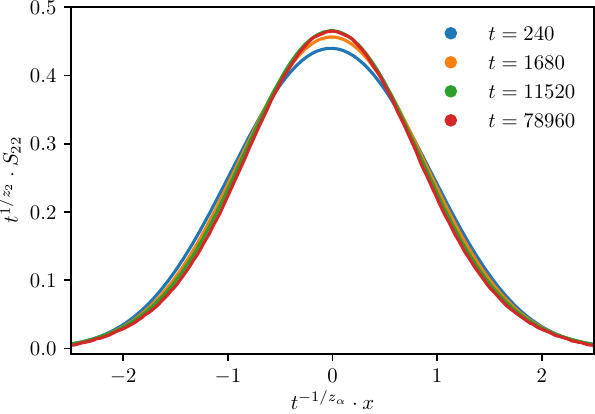}

\caption{Left column: Rescaled $S_{11},S_{22} $ versus rescaled space, for different times. If the hypothesis $z_1= z_2=3/2$ was valid the data should show data collapse. Right column: The best data collapse for large times is achieved with 
$z_1=1.456$, $z_2= 1.587$.
System parameters as in Fig.~\ref{fig:unrotated_correlation}. 
}
\label{FigDataCollapse}
\end{figure}
We next consider the scaling of the correlators in the diagonal basis.
Fig.~\ref{FigZ(t)} shows the main quantities of interest for quantifying the scaling \eqref{scalingform}, the estimated dynamical exponents $z_1(t), z_2(t)$ as functions of
time; according to the scaling hypothesis they should converge to  stationary values, $\lim_{t\to \infty} z_\alpha(t) = z_\alpha$.  
Both the data collapse  of $S_{\alpha\alpha}(x,t)$ at different times, see right column of Fig.~\ref{FigDataCollapse},  and the late time behavior of $z_\alpha(t)$ suggest the validity of the scaling (\ref{scalingform}), with $z_\alpha$
 \begin{align}
&z_1 = 1.456 \pm 0.003 \nonumber\\
&z_2 = 1.587 \pm 0.005 \label{res:z1z2}
\end{align}
where the $z_\alpha$  are $0.5\%$ accurate.  Most notably, both 
$z_1$ and $z_2$ are distinct from the KPZ exponent $z_{\rm KPZ}= 3/2$,  suggested in  \cite{2024Spohn}. 
To emphasize our point, in the left column of Fig.~\ref{FigDataCollapse} we also plot our data for $S_{11},S_{22} $ in a form which should show data collapse if the hypothesis $z_1= z_2= 3/2$ was valid.  Clearly,  the data collapse in the left column of Fig.~\ref{FigDataCollapse} is poor, invalidating the hypothesis. By contrast, the right column of
Fig.~\ref{FigDataCollapse} shows asymptotic data collapse 
for the values $z_1,z_2$ in Eq.~\eqref{res:z1z2}.

\begin{figure}[h]
\includegraphics[width=0.9\columnwidth]{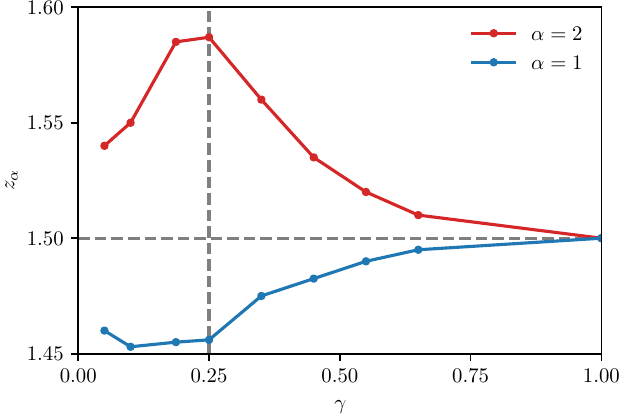}
\caption{Dynamical exponents $z_1, z_2$ versus $\gamma$, obtained from
MC calculations.  The region $\gamma > \gamma_{\rm crit}$ ($\gamma < \gamma_{\rm crit}$) corresponds
to non-isolated UP (isolated UP).  Dynamical exponents are estimated from collapse plots. Systems are of size $L=10^5$ up to $L=5 \times 10^5$. The vertical dashed line  
indicates $\gamma_{\rm crit}$
}
\label{FigZ1Z2vGa}
\end{figure}

\subsection{Other values of the coupling}
To support our findings, we have also investigated the space-time correlator $S_{\alpha\alpha}(x,t)$ at different values of the coupling parameter $\gamma$.  Only for  $\gamma=1$,  the point where the dynamics splits into two independent totally asymmetric processes, do we retrieve the well-known KPZ scaling of a stochastic scalar Burgers equation $z_1=z_2 = 3/2$,  with $ f_{\alpha} = f_{\rm KPZ} $ the celebrated KPZ scaling function obtained in Ref.~\cite{Prae04}.  For all other 
value of the coupling parameter $\gamma<1$ we find not only scaling functions distinct from $f_{\rm KPZ}$ but also different dynamical  exponents, see Fig.~\ref{FigZ1Z2vGa}. 
 We observe a clear tendency of both $z_1,z_2 \rightarrow 3/2 $ as $\gamma$ increases towards $\gamma=1$, 
the point of decoupling into separate simple  exclusion processes. 
Also the shape of scaling functions $f_1,f_2$  become closer to KPZ as $\gamma$ increases towards $1$, data not shown. 
The data for the regime $\gamma < \gamma_{\rm crit}$ which correspond to an isolated umbilic point ( saddle point topology) are less transparent. 
The critical value of the coupling $\gamma = 1/4$, at which the saddle 
point appears,  corresponds to the maximal
$|z_1-z_2|$ difference.
The limit $\gamma \rightarrow 0$  cannot be reached via Monte Carlo simulations since the effective MC dynamics close to $\gamma =0$ at the umbilic point becomes too slow. Further studies of this regime are warranted.  

{
\subsection{Continuum theory}
To substantiate our findings we have also performed numerical simulations of a system of coupled Burgers equations \cite{2024Spohn} for the continuum fields $\phi_\alpha(x, t)$ with $\alpha = 1, 2$
\begin{align}
&\partial_t \phi_1 = \partial_x ( 2b  \phi_1 \phi_2 + D \partial_x \phi_1  + B \xi_1), \label{Burg1}\\
&\partial_t \phi_2= \partial_x ( b  \phi_1^2 +  b \lambda \phi_2^2 + D \partial_x \phi_2  + B \xi_2), \label{Burg2}
\end{align}
where $\xi_\alpha$ are uncorrelated white noise terms with unit variance. The connection with the lattice model's continuum theory is established by the identification $\lambda = 2-\gamma^{-1/2}$, see \ref{app:A}. 
We consider the diagonal dynamical correlators of the continuum model
 \begin{align}
S_{\alpha \alpha}(x,t) = \langle \phi_{\alpha}(x, t) \phi_{\alpha}(0, 0) \rangle. \label{Burgers_corr}
\end{align}
starting from uncorrelated initial states $\langle \phi_\alpha(x, 0), \phi_\beta(x', 0)\rangle=\delta_{\alpha \beta} \delta(x-x')$ at the critical value $\lambda_{\rm crit} = 0$ which corresponds to $\gamma_{\rm crit}=1/4$.

\begin{figure}[h]
\centering
\includegraphics[width=0.98\columnwidth]{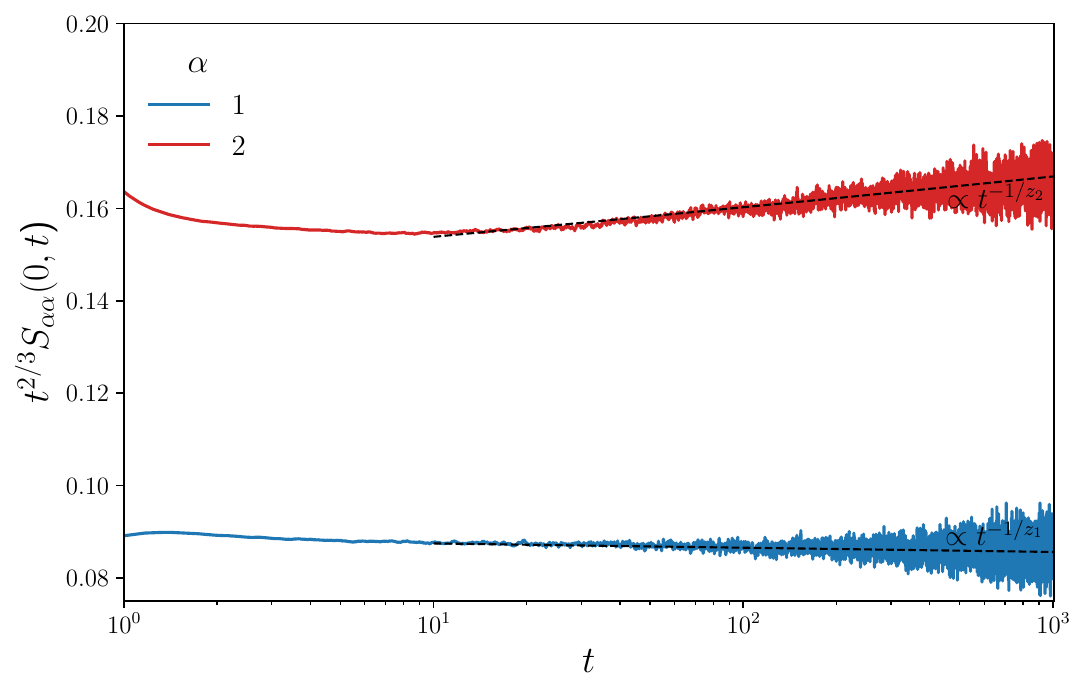}
\caption{{Time-scaling of the diagonal correlators $S_{\alpha \alpha}(0, t)$  of the system of coupled Burgers equations \eqref{Burg1} and \eqref{Burg2}. The data is scaled along the vertical axis such that a dynamical exponent $z_{\rm KPZ}=3/2$ would correspond to asymptotically horizontal data. Dashed black lines indicate asymptotic power-laws with estimated exponents $z_1 \approx 1.49$ and $z_2 \approx 1.54$.
Parameters: number of disretization points=$2^{20}$, $t_{\rm max}=2^{10}$, $b=3$, $\lambda=0$, $D=1$, $B=\sqrt{2}$.}}
\label{fig:Burgers_z}
\end{figure}

We find the most robust signal of the dynamical exponent by considering the time-scaling of the correlator \eqref{Burgers_corr} at $x=0$. As shown in Fig.~\ref{fig:Burgers_z}, we again find clear deviations from the KPZ dynamical exponent $z_{\rm KPZ}=3/2$ for both modes. In particular, we estimate the values of the dynamical exponents as $z_1 \approx 1.49$ and $z_2 \approx 1.54$. While these values for the continuum theory do not precisely match the exponent estimates for the lattice model \eqref{res:z1z2}, we note that the deviations are in the same direction, i.e.~$z_1 < z_{\rm KPZ}$ while $z_2 > z_{\rm KPZ}$. We attribute the quantitative mismatch of the estimates to finite simulation times and different rates of convergence towards stationary values in the lattice and continuum models.
}

\section{Conclusions}
We have presented a numerical study of the dynamical structure factor for a many-body particle system with an umbilic point and two conservation laws.  Both isolated and non-isolated
umbilic point were considered. 
We have investigated the validity of the scaling hypothesis  \eqref{scalingform} in a basis which diagonalizes
 both the dynamical structure matrix $S_{\al \be}(x,t)$ and the static covariance matrix.
 On the basis of the data we conclude that, within the hypothesis \eqref{scalingform} that: (1) the dynamical exponents $z_1, z_2$ and (2) the dynamical scaling functions $f_1,f_2$ vary continuously with the parameter of the model $\ga$ in a nontrivial way, i.e. the scaling functions  are not merely dilated
(the latter observation being  in agreement with an earlier study  \cite{2024Spohn}). {At the umbilic point of the system of coupled Burgers equations we also find systematic deviations from the KPZ dynamical exponents for both modes. 
Continuously varying dynamical exponents $z_1, z_2$, even though unexpected in growth processes,  do not contradict the usual power counting scaling 
arguments (valid for one-component system),  since our two-lane process is not Galilean-invariant, see \cite{1992Kardar,2025Spohn} and also \ref{app:A} for more details. 
}


While continuously varying critical exponents occur in critical statistical model with short range or long range interactions a equilibrium \cite{BaxterBook}, \cite{2025XYlongRangeCriticalExponents}, it is surprising for nonequilibrium stochastic models in one space dimension, such as coupled stochastic multi-component Burgers equations, which can be related to surface growth processes \cite{1997ReviewKrug, Spoh14b}.
In particular,  the mode-coupling theory for multi-component stochastic Burgers equation with 
different characteristic velocities predicts a 
discrete set of dynamical exponents\cite{2015Fibonacci,JSP2015}. 
Our results suggest a need for the extension of mode-coupling theory to systems with an umbilic point and for a further testing of the scaling hypothesis.  

\paragraph*{\bf Acknowledgements}
\v{Z}.K. is supported by the Simons Foundation as a Junior Fellow of the Simons Society of Fellows (1141511). 
V.P.  acknowledges support by ERC Advanced grant
  No.~101096208 -- QUEST, and 
  Deutsche Forschungsgemeinschaft through DFG project
  KL645/20-2. J.S. and V.P. thank G. Sch\"utz for a discussion. \v{Z}.K. thanks Herbert Spohn for useful discussions. We thank the authors of Ref.~\cite{2024Spohn} for providing their data for the comparison in Figs. D1,D2 of \ref{app:D}.\\

\appendix

\section{\\Nonlinear fluctuating hydrodynamics}
\label{app:A}

This Appendix is included for self-containedness. More 
details can be found in \cite{2024Spohn}.
The dynamics of our many-body stochastic process  is a Markov process on an infinite lattice
defined via a classical Master equation with rates  depicted in Fig.~\ref{Fig_BidirectionalModel}.  The particle 
do not change lanes and 
obey an exclusion rule,  i.e. any site can be occupied by at most one particle.  
The process has two conservation laws, the density of particles $u,v$ on lanes $1,2$,   with ranges 
 $0<u,v <1$ because of the exclusion rule.  Steady states of the process  are parametrized by $u,v$,  and are factorized over pairs of vertical lattice sites.  
Namely,  denoting by $\eta_k^\alpha$ a particle number operator on lane $\alpha$ at site $k$,  
the steady state equal time space correlation satisfy 
\begin{align}
\langle \eta_k^\alpha \eta_{k'}^\beta \rangle - \langle \eta_k^\alpha \rangle \langle  \eta_{k'}^\beta \rangle=0,  \  \mbox{if } \ k \neq k'. 
\end{align}
At the same site $k$ 
\begin{align}
&\langle \eta_k^\alpha \eta_{k}^\beta \rangle - \langle \eta_k^\alpha \rangle \langle  \eta_{k}^\beta \rangle=C_{\alpha \beta}\\
&C=
\left(
\begin{array}{cc}
u(1-u)& -u v+ \Omega(u,v)\\
 -u v+ \Omega(u,v)&v(1-v) 
\end{array}
\right)\\
& \Omega(u,v) = \frac{
-1-w_{uv}+\sqrt{(1+w_{uv} )^2 + 4 (\gamma-1) u v}}{2 (\gamma-1)} 
\\
&w_{uv}= (\gamma-1) (1-u-v) 
\end{align}
The steady state currents of the particles on lanes $1,2$ can also be calculated exactly,  yielding 
\cite{Umbilic2012}

\begin{align}
&j_1(u,v) = u(1-u) +w_{uv}\Omega  + (\gamma-1)\Omega^2, \\
&j_2(u,v) = - j_1(v,u).
\end{align}

In the framework of the  nonlinear fluctuating hydrodynamics the time-dependent two-point function at large space $x=a k$ ($a$ being the lattice spacing)  and time 
$\langle \eta_0^\alpha(0) \eta_{x}^\beta(t) \rangle$
(the main quantity of our interest for the scaling analysis) is governed by the first and the second derivatives of the steady currents 
$j_\alpha$ at a given point $u,v$.  

Namely,  denoting the local density of particles at lane $\alpha$ as $\rho_\alpha(x,t)$ and the average densities as $\rho_\alpha$,
i.e.  $\rho_1 \equiv u$,  $\rho_2 \equiv v$,
one writes a system of nonlinear fluctuating hydrodynamics equations  for the density fluctuations 
$\varphi_\alpha (x,t) = \rho_\alpha(x,t) - \rho_\alpha$
as 
\begin{align}
&\partial \vec{\varphi} = - \partial_x \left( J \vec{\varphi} + \frac12 \langle{\varphi} | \vec{H} |{\varphi}\rangle - \partial_x \tilde D \vec{\varphi} + \tilde B \vec{\tilde \xi}
\right) \label{eq:NLFH}
\end{align}

Here $J$ and $\vec{\varphi} = (\varphi_1,\varphi_2)^T$,  $\vec{H} = (H_1,H_2)^T$, $J$ is the Jacobian 
$J_{\alpha \beta} = \partial j_\alpha/\partial \rho_\beta$
 and $H^\alpha$ is the Hessian,  $(H^\alpha)_{\beta \gamma} = \partial^2 j_\alpha/(\partial \rho_\beta \partial \rho_\gamma) $.
$\tilde B \vec{\tilde \xi}$ describes a noise and $\tilde D$
is  a phenomenological diffusion matrix.

At the  umbilic point (UP) $\rho_1=\rho_2=1/2$ we have 
\begin{align}
&J=0 \label{app:J=0}\\
&H_1 =\left(
\begin{array}{cc}
 \frac{1-\gamma }{2\gamma^{1/2}}-2 & -\frac{\left(\gamma^{1/2}-1\right)^2}{2 \gamma^{1/2}} \\
 -\frac{\left(\gamma^{1/2}-1\right)^2}{2 \gamma^{1/2}} &
   \frac{1-\gamma }{2 \gamma^{1/2}} \\
\end{array}
\right)\\
&H_2=-H_1 - 2 \sigma^z\\
&C =\left(
\begin{array}{cc}
\frac{1}{4}& -\frac{1}{4} + \Omega_{UP}\\
-\frac{1}{4} + \Omega_{UP} &\frac{1}{4}
\end{array}
\right) \\
&\Omega_{UP}\equiv  \Omega\left(\frac{1}{2},\frac{1}{2}\right)=\frac{1}{2 + 2 \gamma^{1/2}},
\end{align} 

Now,  we shall choose a transformation $R$ such that $(R C R^T)$ is the $2\times 2$ unit matrix:.
\begin{align}
&R =
\left(
\begin{array}{cc}
\frac{1}{\sqrt{2 d_1}}& 0\\
0&\frac{1}{\sqrt{2 d_2}}
\end{array}
\right)   \ 
\left(
\begin{array}{cc}
1& 1\\
-1&1 
\end{array}
\right),   \label{app:R} \\
&d_1 = \Omega_{UP},
\quad d_2 =\frac{1}{2}- \Omega_{UP}\\
&(R C R^T)_{\alpha \beta} = \delta_{\alpha \beta} \label{app:C=I}
\end{align}
In terms of  new variables $\vec{\phi} = R \vec{\varphi}$ at UP Eq.~(\ref{eq:NLFH}) accounting for 
(\ref{app:J=0}) is rewritten as 
 
\begin{align}
&\partial_t \vec{\phi} = - \partial_x \left(\langle{\phi} |\vec{G}|{\phi}\rangle - \partial_x D \vec{\phi} + B \vec{\xi}
\right) \label{eq:NLFH1}
\end{align}
where $D=R \tilde{D} R^T$,   $\vec{G} =(G^1,G^2)^T$ with $G^\alpha$  conventionally called mode coupling matrices,  related to $H_\alpha$ via 
\begin{align}
&G^\alpha= \frac12 \sum_\beta R_{\alpha \beta} (R^{-1})^T H^\beta R^{-1},
\end{align}
Explicitly,   one finds
\begin{align}
&G^1= b \left(
\begin{array}{cc}
0& 1\\
1&0
\end{array}
\right)\\
&G^2= b \left(
\begin{array}{cc}
1& 0\\
0&2-\gamma^{-1/2}
\end{array}
\right)\\
&b= \frac{1}{2} \sqrt{\frac{\gamma^{1/2}}{\gamma^{1/2}+1}}
\end{align}
Assuming diagonal difusion matrices $D_{\alpha \beta} = D_\alpha \delta_{\alpha \beta}$ and 
writing the system (\ref{eq:NLFH1})  in components,  we have
\begin{align}
&\partial_t \phi_1 = \partial_x ( 2b  \phi_1 \phi_2 + D_1 \partial_x \phi_1  + B_1 \xi_1) \label{app:eq1}\\
&\partial_t \phi_2= \partial_x ( b  \phi_1^2 +  b\left(2-\gamma^{-1/2}\right) \phi_2^2 + D_2 \partial_x \phi_2  + B_2 \xi_2) \label{app:eq2}\\
& \langle \phi_\alpha(x), \phi_\beta(x')\rangle=\delta_{\alpha \beta} \delta(x-x') \label{eq:covarianceStatic}
\end{align}
where (\ref{eq:covarianceStatic}) follows from (\ref{app:C=I}). 
The above system of equations,  up to phenomenological constants of  diffusion and noise terms,  coincides with the  
system of stochastic Burgers equations studied by Spohn and collaborators, see Eq (3.1) in  \cite{2024Spohn}. \\

{
Note that the system of Eqs.~(\ref{app:eq1},\ref{app:eq2}) is not invariant under the Galilean transformation
 \begin{align}
& \phi_2(x,t) = B +  \phi_2(x + 2 b Bt ,t) \\
& \phi_1(x,t) =   \phi_1(x + 2 b Bt ,t),
\label{app:GI}
\end{align}
except
 at the point  $2-\gamma^{-1/2}=1$,  or $\ga=1$ i.e.   at the point where the system splits into the two independent one-lane exclusion 
processes,  governed by the KPZ universality,  see Fig.~\ref{Fig_BidirectionalModel} and Fig.~\ref{FigZ1Z2vGa}.  
Away from the point $\ga=1$ the absence of Galilean invariance, as argued in \cite{1992Kardar, 1989Kardar,2025Spohn},  breaks the validity of the 
scaling identity $\xi_2+z_2 =2$ between the roughening exponent 
$\xi_2$ (equal to $\frac12$ ) and the dynamical exponent $z_2$,  yielding the dynamical exponents $z_2$ (and similarly,  $z_1$) free of constraints.  
}\\

To access the space-time  correlation matrix $S_{\alpha \beta} = \langle \phi(0,0) \phi(x,t) \rangle$ 
for the system of stochastic Burgers equations  (\ref{app:crit}),(\ref{app:crit1})
 we perform a Monte-Carlo study of the two-lane particle system of Fig.~\ref{Fig_BidirectionalModel} for $\gamma = \gamma_{crit}=1/4$.  In the Monte-Carlo study we measure the correlation matrix 
$\tilde S_{\alpha \beta} = \langle \varphi(0,0) \varphi(x,t) \rangle$. 
The change of variables 
$\phi_\alpha = \sum_{\beta=1,2}R_{\alpha \beta} \varphi_{\beta}$ with $R$ from (\ref{app:R}) leads to 
    
\begin{align}
S_{11}(x,t)&=\frac{1}{2 d_1} \left(\tilde{S}_{11}(x,t) + \tilde{S}_{22}(x,t) +\tilde{S}_{12}(x,t) + \tilde{S}_{21}(x,t)\right)\\
S_{22}(x,t)&=\frac{1}{2 d_2} \left(\tilde{S}_{11}(x,t) + \tilde{S}_{22}(x,t) -\tilde{S}_{12}(x,t) - \tilde{S}_{21}(x,t)\right) \label{app:Sab}\\
&\frac{1}{2 d_1} = (1+\gamma^{1/2}), \quad \frac{1}{2 d_2} = (1+\gamma^{-1/2})
\end{align}
reported in the main text, see  (\ref{eq:StildeS11}),  (\ref{eq:Sab}).
For $\gamma = \gamma_{crit}=1/4$, the above system acquires an especially simple form 
\begin{align}
&\partial_t \phi_1 = \partial_x ( 2b  \phi_1 \phi_2 + D_1 \partial_x \phi_1  + B_1 \xi_1)  \label{app:crit}\\
&\partial_t \phi_2= \partial_x ( b  \phi_1^2  + D_2 \partial_x \phi_2  + B_2 \xi_2) \label{app:crit1}
\end{align}

\section{\\Simulation method for two-point functions}
\label{app:B}

Initial states are drawn from the stationary distribution of the process. No relaxation is required.
The two-point function can be estimated using translation invariance and stationarity, which allow for the computation of the spatial and temporal averages. To account for computationally expensive pseudo random number generation we generate $R$ independent initial states and propagate them with the same set of random numbers, leading to the Monte-Carlo estimator
\begin{align}
\tilde\sigma_{x,t}^{\lambda \mu}(M,\tau,L,R) =&
\frac{1}{LMR} \sum_{l=1}^L \sum_{j=1}^M \sum_{r=1}^R
n_{l+x, j\tau+t}^{\lambda,(r)}n_{l,j\tau}^{\mu,(r)}.
\label{eq-sigma}    
\end{align}
Finally, $\tilde S_{\lambda \mu}(x,t)$ is obtained by the average over
$P$ independently generated and propagated initial configurations
of $\tilde\sigma_{L,x}^{\lambda \mu}$, i.e.
\begin{align}
\tilde{S}_{\lambda\mu}(x,t)=
\frac{1}{P}\sum_{p=1}^P \tilde\sigma_{L,x}^{\lambda \mu, (p)}
- \rho_\lambda\rho_\mu
+\mathcal{O}(P^{-1/2}).
\end{align}
The error estimates for
$S_{x}^{\lambda \mu}(t)$ are calculated from $P$ independent measurements, whereas $L$, $M$, $\tau$ and $R$ are variance reduction parameter.

\section{\\Time resolved dynamical exponent}
\label{app:C}

To compute the dynamical exponent as a function of time, we first smooth the data by determining a differentiable spline function $s_\alpha(\tau)$ by minimizing the functional
\begin{align}
F_p \left[\{S_{\alpha \alpha}(0, t)\}_{t=0}^{t_{\rm max}}\right] = 
p\sum_{t=1}^{t_{max}}
\frac{\left|t^{2/3}S_{\alpha\alpha}(0,t)-s_{\alpha}(\ln t) \right|^2}
{t\mathrm{Var}\left(t^{2/3}S_{\alpha\alpha}(0,t)\right)/P}\nonumber\\
+(1-p)\int_{0}^{\ln t_{max}}\left | \frac{\mathrm d^2 s_\alpha}{\mathrm d \tau^2}\right|^2 \mathrm{d}\tau.
\end{align}
The parameter $p \in [0, 1]$ determines the trade-off between errors and roughness of fit. We determine the optimal choice of $p$ by decreasing $p$ and stopping when the solution becomes unstable, while errors are estimated and controlled by bootstrapping techniques.
The resulting  smoothed fit function $s_\alpha(\tau)$ is related to the structure function as
\begin{align}
S_{\alpha\alpha}(x=0,t) \simeq t^{-2/3} s_\alpha(\ln t) .
\end{align}
The time-resolved dynamical exponent
\begin{align}
z_\alpha(t)=\left(-\frac{\mathrm d \ln S_{\alpha\alpha}(0,t)}{\mathrm d \ln t}\right)^{-1}
\end{align}
is expressed in terms of the smoothed fit functions as
\begin{align}
z_\alpha(t) \simeq \left(\frac{2}{3} - \frac{1}{s_\alpha(\tau)}\left.\frac{\mathrm{d}s_\alpha}{\mathrm{d}\tau}\right|_{\ln t}\right)^{-1}.
\end{align}

{
\section{\\Comparison of  intermediate-time data with the data of  Ref.~\cite{2024Spohn}}
\label{app:D}
In this Appendix we compare our data against the data of  Figs.1,4  of Ref.~\cite{2024Spohn}.  In Figs.~\ref{Fig_comparison_lattice} and \ref{Fig_comparison_cont} we show the comparisons of space-time correlators of the two-lane lattice model and of the system of coupled Burgers equations respectively. In both cases our data (shown by colored lines) agree with the data of Ref.~\cite{2024Spohn} at comparable times. We note that while the dynamical exponent $z_{\rm KPZ}=3/2$ appears to fit the data rather well (only the off-diagonal correlator in the lattice models fails to collapse fully), this is only due to the insufficiently short time window (the maximal and the minimal times  in  Fig.1 and in Fig.4 of Ref.~\cite{2024Spohn} differ just by a factor of 2 ). To observe the  deviations from $z_{\rm KPZ}=3/2$ a sufficiently long time window is needed, e.g. in our Fig.~\ref{FigDataCollapse}
we have $t_{\max}/t_{\min} > 300 )$.

\begin{figure}[htb]
\includegraphics[width=0.9\columnwidth]{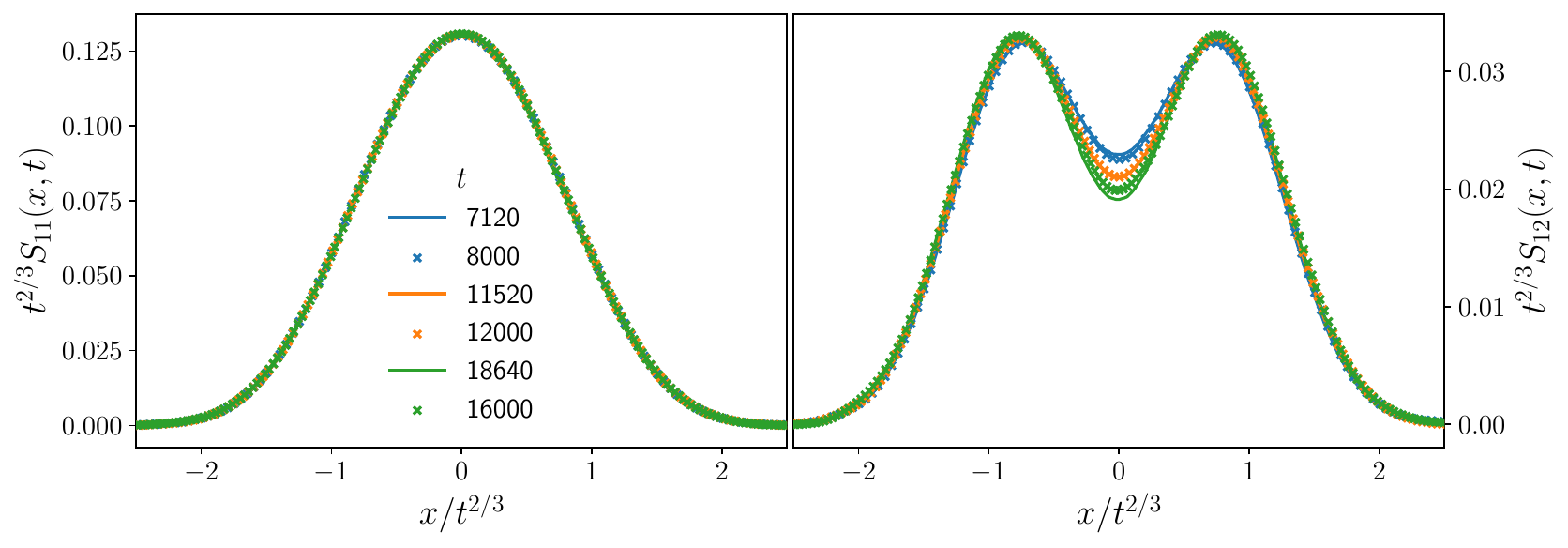}
\caption{Comparison of structure factors of the two-lane Markov process (colored lines) scaled with dynamical exponent $z=3/2$ against data presented in Figure 4 of Ref.~\cite{2024Spohn} at comparable times (colored crosses) .
Simulation parameters as in Fig.~\ref{fig:unrotated_correlation}.
}
\label{Fig_comparison_lattice}
\end{figure}

\begin{figure}[htb]
\includegraphics[width=0.9\columnwidth]{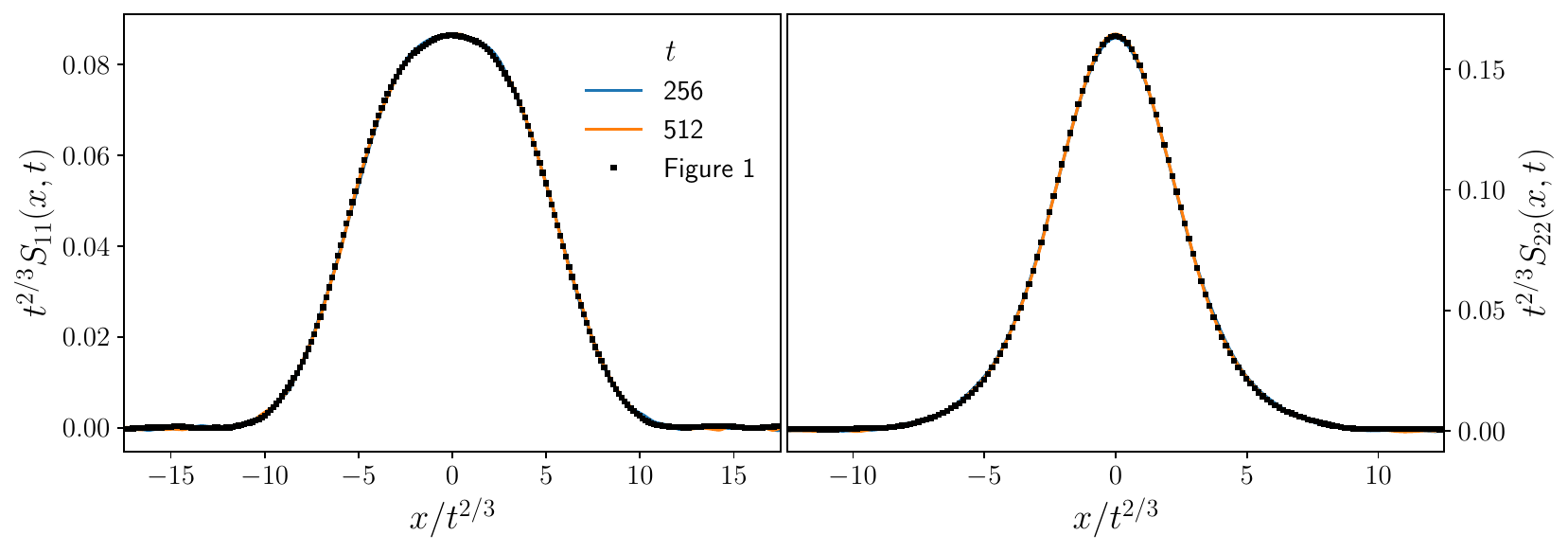}
\caption{Comparison of structure factors of the system of coupled Burgers equations \eqref{Burg1},\eqref{Burg2} (colored lines) scaled with dynamical exponent $z_{\rm KPZ}=3/2$ against data in Figure 1 of Ref.~\cite{2024Spohn} at comparable times (black squares, averaged over all three times $t=200, 300, 400$) .
Simulation parameters as in Fig.~\ref{fig:Burgers_z}.}
\label{Fig_comparison_cont}
\end{figure}

}

\end{document}